\documentclass[a4paper]{article} 

\def\journalfont{\rm}         
\def\jou#1{{\journalfont #1\ }}
\def\joudef#1#2{\def #1{\jou{\ignorespaces #2}}}

\joudef{\aaa}    { Astron.\ Astrophys.}
\joudef{\aip}    { Adv.\ Phys.}
\joudef{\adm}    { adv.\ math.}
\joudef{\am}     { Ann.\ Math.}
\joudef{\apl}    { Ann.\ Phys.\ (Leipzig)}
\joudef{\apny}   { Ann.\ Phys.\ (N.Y.)}
\joudef{\arnps}  { Annu.\ Rev.\ Nucl.\ Part.\ Sci.}
\joudef{\apj}    { Astrophys.\ J.}
\joudef{\apjl}   { Astrophys.\ J.\ Lett.}
\joudef{\cjp}    { Can.\ J.\ Phys.}
\joudef{\cmp}    { Commun.\ Math.\ Phys.}
\joudef{\cqg}    { Class.\ Quantum Grav.}
\joudef{\grg}    { Gen.\ Rel.\ Grav.}
\joudef{\ijmpd}  { Int.\ J.\ Mod.\ Phys.\ D}
\joudef{\ijtp}   { Int.\ J.\ Theor.\ Phys.}
\joudef{\invm}   { Invent.\ Math.}
\joudef{\jm}     { J.\ Math.}
\joudef{\jmaa}   { J.\ Math.\ Anal.\ Appl.}
\joudef{\jmp}    { J.\ Math.\ Phys.}
\joudef{\jpa}    { J.\ Phys.\ A}
\joudef{\lr}     { Liv.\ Rev.\ Rel.}
\joudef{\mnras}  { Mon.\ Not.\ R.\ Ast.\ Soc.}
\joudef{\mpl}    { Mod.\ Phys.\ Lett.}
\joudef{\mpla}   { Mod.\ Phys.\ Lett.\ A}
\joudef{\nature} { Nature}
\joudef{\nc}     { Nuovo Cim.}
\joudef{\npb}    { Nuc.\ Phys.\ B}
\joudef{\ph}     { Physica}
\joudef{\pla}    { Phys.\ Lett. A}
\joudef{\plb}    { Phys.\ Lett. B}
\joudef{\pr}     { Phys.\ Rev.}
\joudef{\pra}    { Phys.\ Rev.\ A}
\joudef{\prb}    { Phys.\ Rev.\ B}
\joudef{\prc}    { Phys.\ Rev.\ C}
\joudef{\prd}    { Phys.\ Rev.\ D}
\joudef{\prep}   { Phys.\ Rep.}
\joudef{\prl}    { Phys.\ Rev.\ Lett.}
\joudef{\prsla}  { Proc.\ Roy.\ Soc.\ Lond.\ A}
\joudef{\ptp}    { Prog.\ Theor.\ Phys.}
\joudef{\ptps}   { Prog.\ Theor.\ Phys.\ Suppl.}
\joudef\rmp      { Rev.\ Mod.\ Phys.}
\joudef\spj      { Sov.\ Phys.\ JETP}

\newcommand{\eprint}{{\it Preprint:} \textsf} 
\newcommand{\email}{\textsf}

\newcommand\be{\begin{equation}} \newcommand\ee{\end{equation}} 
\newcommand\bd{\begin{displaymath}}\newcommand\ed{\end{displaymath}}

\renewcommand{\d}{{\rm d}} 

\newcommand\ts\textstyle

\def\eg{{\it e.g.}} \def\etal{{\it et al.}} \def\ie{{\it i.e.}}
\def\cf{{\it cf.}}

\newtoks\reportnoregister \newtoks\eprintnoregister
\newcommand{\reportnumber}[1]{\reportnoregister={#1}}
\newcommand{\eprintnumber}[1]{\eprintnoregister={#1}}

\reportnumber{\mbox{}} 
\eprintnumber{\mbox{}} 

\newcommand{\reportid}{
   \begin{minipage}{17cm}\vspace{-3.2cm}
     \begin{flushright}
      {\normalsize \the\reportnoregister \\[-.2cm]
            \email{\the\eprintnoregister}}\vspace{3.2cm}
     \end{flushright}
   \end{minipage}\hspace{-17cm} }

\catcode`@=11   
\def\title#1{\gdef\@title{\reportid#1}}
\catcode`@=12   

\newcommand{\ncd}{\newcommand}
\ncd{\beq} {\begin{equation}}
\ncd{\eeq} {\end{equation}}
\ncd{\BE} {\begin{eqnarray}}
\ncd{\EE} {\end{eqnarray}}
\ncd{\lbeq}[1]  {\label{eq: #1}}
\ncd{\refeq}[1] {(\ref{eq: #1})}
\ncd{\mrm}    {\mathrm}
\ncd{\ms}{\mathstyle}
\ncd{\ds}{\displaystyle}
\ncd{\der}{{\mathrm{d}}}
\ncd{\rtil}{\tilde{r}}
\ncd{\Mr}{\frac{2M}{r}}
\ncd{\rhotil}{\tilde{\rho}}
\ncd{\rstar}{r_{*}}

\usepackage{amsmath}
\usepackage{amssymb}

\ncd{\dell}{\partial}
\ncd{\mnote}[1]{\marginpar{\small #1}}

\ncd{\ec}{\check\epsilon}
\ncd{\rhoc}{\check\rho}
\ncd{\muc}{\check\mu}
\ncd{\pc}{\check{p}}
\ncd{\Gc}{\check\Gamma}
\ncd{\Oc}{\check\Omega}
\ncd{\betac}{\check\beta}

\ncd{\bldeta}{\boldsymbol{\eta}}
\ncd{\bldone}{\mathbf{1}}
\ncd{\blds}{\mathbf{s}}
\ncd{\bldk}{\mathbf{k}}
\ncd{\blde}{\mathbf{e}}

\ncd{\abs}[1] {|#1|}
\ncd{\ubold}{\mathbf u}
\ncd{\Abold}{\mathbf A}
\ncd{\Bbold}{\mathbf B}
\ncd{\Mbold}{\mathbf M}
\ncd{\tsfrac}[2]{{\ts\frac{#1}{#2}}}
\ncd{\lagom}{\hspace{.6pt}}
\ncd{\muk}{k}
\ncd{\lagomdot}{{\mbox{\large$\cdot$}}}

\ncd{\stil}{\tilde{s}}
\ncd{\ftil}{\tilde{f}}
\ncd{\Otil}{\tilde{\Omega}}

\ncd{\D}{\mathcal{D}}
\ncd{\Qcal}{\mathcal{Q}}
\ncd{\Jcal}{\mathcal{J}}
\ncd{\Ecal}{\mathcal{E}}
\ncd{\Wcal}{\mathcal{W}}
\ncd{\Xcal}{\mathcal{X}}
\ncd{\Ycal}{\mathcal{Y}}
\ncd{\Bcal}{\mathcal{B}}
\ncd{\Acal}{\mathcal{A}}
\ncd{\Scal}{\mathcal{S}}
\ncd{\Ccal}{\mathcal{C}}

\ncd{\vt}{v_{t\perp t}^{\,2}}
\ncd{\vr}{v_{r\perp}^{\,2}}

\ncd{\psimap}{\Psi}
\ncd{\psivar}{\psi}

\ncd{\Ap}{(r\psivar)'}
\ncd{\Adot}{(r\psivar)^{\lagomdot}}
\ncd{\Bp}{(r^{-1}\varphi)'}
\ncd{\Bdot}{(r^{-1}\varphi)^{\lagomdot}}

\ncd{\ela}{\left(1-\frac{2m}{r}\right)}
\ncd{\nfrac}[2]{\left(\frac{n_{#1}}{n_{#2}}\right)^2}

\ncd{\shm}{S}
\ncd{\shmtwoD}{\mathcal{\shm}}
\ncd{\lie}{\mathcal{L}}
\ncd{\brk}{\mathrm{max}}
\ncd{\fgauge}{f_\mathrm{G}}

\ncd{\I}{\mrm{c}}
\ncd{\f}{\mrm{f}}

\ncd{\tabnl}{\\[-8pt]}

\begin{document}

\title{Axial quasi-normal modes of neutron stars: Accounting for the superfluid in the crust}
\author{Lars Samuelsson$^{1,2}$\footnote{E-mail: \email{larsam@nordita.org}}~~and
        Nils Andersson$^2$\footnote{E-mail: \email{na@maths.soton.ac.uk}} \\[-10pt]
{\small $^1$ NORDITA, AlbaNova University Center, SE-106 91 Stockholm, Sweden} \\[-15pt]
{\small $^2$ School of Mathematics, University of Southampton, Southampton SO17~1BJ, UK} \\
\begin{minipage}[t]{0.8\linewidth}\small{We present the results
of the first study of global oscillations of relativistic stars
with both elastic crusts and interpenetrating superfluid
components. For simplicity, we focus on the axial quasi-normal
modes. Our results demonstrate that the torsional crust modes are
essentially unaffected by the coupling to the gravitational
field. This is as expected since these oscillations are known to be
weak gravitational-wave sources. In contrast, the presence of a
loosely coupled superfluid neutron component in the crust can have a
significant effect on the oscillation spectrum. We show that the
entrainment between the superfluid and the crust nuclei is a key
parameter in the problem. Our analysis highlights the need for a more
detailed understanding of the coupled crust-superfluid at the
microphysical level. Our numerical results have, even though we have not
considered magnetised stars, some relevance for efforts to carry out
seismology based on quasi-periodic oscillations observed in the tails
of magnetar flares. In particular, we argue that the sensitive
dependence on the entrainment may have to be accounted for in attempts
to match theoretical models to observational data.
}
\end{minipage}}

\maketitle

\section{Introduction}

Following the observations of quasi-periodic oscillations (QPOs) in
the tails of giant flares in several soft gamma-ray repeaters (SGRs)
\cite{sw:qpo,sw:flare2,israel:qpo}, axial (torsional) oscillations
of neutrons stars have attracted considerable interest. The SGRs are
generally thought to be highly magnetised neutron stars (magnetars)
\cite{dt:magnetars}, and it would be natural to explain
QPOs in the range $\sim 30 - 150$~Hz as axial crustal
oscillations. That magnetar activity might trigger such oscillations
was, in fact, suggested quite some time ago \cite{duncan:1998A}. This
explanation is supported by the spacing of the QPO frequencies
\cite{sa:axicowling}. The model does, however, face a serious
challenge since the strong magnetic field should couple the crust to
the core in less than an oscillation period
\cite{levin:magnetars,gsa:mhd}. In addition, the magnetic field will
affect the motion of the crust itself
\cite{piro:flares,sks:torsional,sksv:torsionalII}. To determine global
``elasto-magnetic'' oscillation modes is challenging due to the
possible existence of an Alfv\'en continuum in the core. The presence
of this continuum casts doubt on the very existence of global mode
solutions with a discrete frequency spectrum, see
\cite{cbk:QPOs,csf:QPOs} for recent progress on the purely fluid problem.
However, recently there has been some evidence
\cite{levin:magnetarsII,lee:aximag} in favour of the hypothesis put
forward in \cite{gsa:mhd}. For moderate magnetic fields ($B\lesssim
10^{15}$ G) the global modes most easily excited by a catastrophic
event in the crust have frequencies that are tuned to the purely elastic
crustal mode frequencies. For very high magnetic fields $\sim
10^{16}$~G the situation is not quite as clear \cite{lee:aximagII}.

Most available studies of the axial mode problem have completely
ignored the dynamical role of the dripped neutrons in the inner
crust. Since these neutrons are likely to be superfluid they can flow
through the crust lattice. This leads to the problem having an
additional degree of freedom, and it is important to establish to what
extent the presence of this new component will affect the
seismology. A recent local analysis \cite{ags:supsig} has shown that
the superfluid components, both in the crust and the core, may have
significant effects. The main aim of this paper is to consider this
problem in more detail. We will, for the first time, examine the
effects of the superfluid crust component on the global axial
oscillation spectrum. The obtained results have immediate relevance for the
magnetar discussion. We are also reporting real progress in modelling
the dynamics of realistic neutron stars. Our discussion provides
insight into the global dynamics of solids coexisting with a
superfluid component, and highlights the microphysics parameters that
are needed if we want to improve the analysis. As we will see, the
entrainment between the superfluid neutrons and the crust nuclei (the
``effective mass'' of the free neutrons) is a key parameter that needs
to be constrained better by equation of state calculations.

Our analysis is based on state-of-the-art matter modelling. We employ
a linear perturbation scheme in general relativity, and do not make
additional approximations such as the Cowling approximation, wherein
the gravitational degrees of freedom are neglected (or, as discussed
in \cite{sa:axicowling}, the coupling is ignored). Since we account
for the dynamics of spacetime itself, the oscillation modes we
determine can be split into two distinct families, the torsional
$t$-modes associated with the matter motion supported by the elastic
properties of the solid, and the gravitational $w$-modes (which in
turn can be subdivided into different classes, see \eg\
\cite{ks:qnm}). In principle, we should be able to address directly
the gravitational-wave damping of the $t$-modes and the influence of
the elastic and superfluid nature of the matter on the $w$-modes. In
practice, however, we are unable to compute the damping time of the
$t$-modes. This result is obvious if we estimate the coupling between
the torsional motion of a solid and the gravitational field. Using the
quadrupole formula, Schumaker \& Thorne \cite{st:torsional} estimate
the gravitational-wave damping time scale of the $t$-modes to be $\sim
10^{4}$ years, \ie\ some fourteen orders of magnitude longer than the
typical oscillation time-scale for a 30~Hz mode. The upshot of this is
that the complex angular frequency $\omega$ in the standard ansatz for
the time-dependence $e^{i\omega t}$ has an imaginary part which is
fourteen orders of magnitude smaller than the real part. This is a
problem for any numerical scheme that aims at determining the complex
$\omega$. The, perhaps naive, numerical scheme we use here is
certainly unable to compute the imaginary part of these very slowly
damped modes. We are, however, able to compute the real part, and
hence the frequencies, of the $t$-modes accurately. The obtained
results confirm the expectation that these modes can be accurately
determined within the Cowling approximation \cite{sa:axicowling}. This
is true regardless of whether the superfluid neutrons are taken into
account or not. The free neutrons on the other hand \emph{do} affect
the axial mode frequencies, in agreement with the qualitative results
in
\cite{ags:supsig}. The effect on the global mode frequencies turns out
to be at the 10\% level, but depends strongly on the precise
properties of the inner crust. Overall, our results imply that, as far
as seismology efforts are concerned \cite{sa:axicowling}, the Cowling
approximation should be sufficient. We will return to that problem,
and detailed implications of the presence of the free neutrons,
elsewhere.

The $w$-modes pose no real technical problems for our numerics and we
can determine their frequencies and damping times to about 10 digit
precision. The results indicate that the inclusion of a more accurate
treatment of the matter content in the star does not influence the
$w$-mode spectrum beyond the $\sim 10^{-8}$ level.  This could have
been anticipated since the $w$-modes depend almost entirely on the
gravitational field of the background star. The background model is
hardly affected at all by a more refined matter modelling, \eg\ in the
crust region.

The plan of the paper is as follows. In section \ref{sec:formalism} we
briefly review the formalism needed for a general relativistic
treatment of the dynamics of solids coexisting with a
(super)fluid. This is followed, in section \ref{sec:perturbations}, by
an explanation of how the free neutrons affect the equations of motion
in the linear perturbation regime. Our numerical results are then
presented in section \ref{sec:results}. We conclude the paper with
section \ref{sec:discussion} and a discussion of the results and
possible future developments.  The numerical machinery used to solve
the equations is summarised in an Appendix. 

\section{Formalism}
\label{sec:formalism}

We begin our analysis by introducing the basic framework used to
describe the dynamics of a superfluid immersed in a solid lattice. The
discussion is based on the theory developed by Carter \& Samuelsson
(CS) \cite{cs:superelastic}, to which we refer for details. We will,
however, adapt our discussion to make maximal use of the formalism
developed in a series of papers by Karlovini \etal\ dealing with pure
solids \cite{ks:relasticityI,ksz:stability,ks:exact,ks:relastaxial},
hereafter referred to as Papers {I}-{IV}. We will also, as much as
possible, adapt the notation to the recent review of relativistic
fluids by Andersson \& Comer \cite{ac:sfreview}, hereafter AC.

The theory is derived from a Lagrangian ``master function'' $\Lambda$
which plays the dual role of providing the matter part of the total
action and also giving the equation of state (EoS). For the purposes
of the present study it is convenient to decompose the Lagrangian as
(see CS)
\beq
  \Lambda = \Lambda_{\mrm{liq}} + \Lambda_{\mrm{sol}}
\eeq
where $\Lambda_{\mrm{liq}}$ is a function of the particle fluxes
$n_\I^a$ and $n_\f^a$, while the elastic properties of the solid
lattice are contained within $\Lambda_{\mrm{sol}}$. The functional
$\Lambda_{\mrm{liq}}$ is the contribution due to the liquid properties
and corresponds to the full Lagrangian for a two-fluid system (see
\eg\ AC). We use the constituent indices '$\I$' and '$\f$' to
distinguish the ``confined'' baryons in the lattice and the ``free''
neutrons, respectively. We assume a metric signature of the form $(-,+,+,+)$ and
use early Latin letters, '$a$', '$b$', '$d$', \ldots\ to denote
abstract spacetime indices, see
\eg\ \cite{wald:gr}, omitting '$c$' in order to avoid confusion.
We will employ liberal index positioning to avoid unnecessary
cluttering of the formulae. For instance, the total free neutron
number density is given by
\beq
  n_\f^2 = -n_\f^an_a^\f = -g_{ab}n_\f^a n_\f ^b
\eeq
and similar for the confined baryon number density $n_\I$.

The liquid contribution can be further
split into a term that is independent of the relative velocity and a piece
describing the effects of the relative current,
\beq
  \Lambda_{\mrm{liq}} = -\rhoc + \Lambda_{\mrm{ent}}
\eeq
Here we denote the velocity independent term by $\rhoc$ since it
corresponds to the comoving unsheared energy density (which is
uniquely defined only for zero relative velocity) as described, \eg\,
in Paper I. This quantity is assumed to describe the minimum energy
for a given total baryon density $n=n_\I+n_\f$ and we will therefore
consider it to be a function of $n$ only. The remaining term arises
because the state of matter may explicitly depend on the relative
velocity between the different species of particles in a multi-fluid
system. This leads to an effect known as entrainment. If only low relative
velocities are considered (as will be the case here) it is sufficient
to consider a slow motion approximation of the entrainment
contribution. Since $\rhoc$ represents a minimum energy state for a
given baryon density $n$ it is clear that the leading order term in
the entrainment term must be quadratic. The question is in what? There are
at least two choices. In CS the expansion was carried out in terms of
the relative flux, $n_\perp^a$, which, geometrically, can be defined
locally as the projection of the neutron current onto the space
orthogonal to the four velocity $u^a$ of the solid (or vice
versa),
\beq\lbeq{nperp}
  n_\perp^a = h^a{}_bn_\f^b, \qquad h_{ab} = g_{ab} + u_a u_b
\eeq
On the other hand, in micro-physical calculations it seems customary
\cite{cj:entrainment,chamel:2fluid} to use the relative velocity $v^a$
(or some proxy of this) which is related to the neutron current
through
\beq\lbeq{nfproj}
  n_\f^a = \gamma n_\f(u^a + v^a),
  \qquad u^av_a = 0,
  \qquad \gamma = (1-v^2)^{-1/2},
  \qquad v^2 = v^av_a
\eeq
Combining \refeq{nperp} and \refeq{nfproj} we see that
\beq
   n_\perp^a = \gamma n_f v^a \quad \Rightarrow \quad n_\perp^2 = n_f^2 \frac{v^2}{1-v^2}
       \approx  n_f^2 v^2(1+v^2)
\eeq
This analysis shows that the two prescriptions agree to
$O(v^2)$. However, we also find that the equations of motion, which
will depend on derivatives of the master function, may differ at
$O(v^2)$ depending on the choice of expansion and the level of
truncation. In order to be consistent, one should therefore be careful
to employ the same scheme as the microphysics calculation on which one
bases the continuum model. This level of consistency is difficult to
reach given our present (rather basic) level of understanding, but it sets
the standard that future work should aspire to.

In this study we will consider axial linear perturbations of a static
background model. The incompressible nature of this kind of motion
together with the vanishing of the relative velocity in the background
implies that we will not, even in principle, need to know the master
function beyond $O(v^2)$. Thus, for the present purposes the problem
mentioned above is a non-issue. For more general backgrounds and
compressible motion (\eg\ associated with acoustic oscillation modes)
caution is advised.

Since the microphysics calculations we take our parameters from are
performed as an expansion in $v$ this is what we will use. Then,
making the standard assumption that the liquid is intrinsically
isotropic, we can write
\cite{cj:entrainment,chamel:2fluid}
\beq
  \Lambda_{\mrm{ent}} = \frac{1}{n_\I}m_\I^\f (n_{\I\f}^2-n_\I n_\f)
      = n_\f m_\I^\f(\gamma-1) = \frac{1}{2}n_\f m_\I^\f v^2 + O(v^4)
\eeq
where $m_\I^\f = m^*_\f - m$, $m^*_\f$ is the effective (dynamical)
mass of the neutrons, $m$ is the neutron mass (which we may take to be
equal to the proton mass) and $n_{\I\f}^2 = -n_\I^a n^\f_a$.

The solid contribution is most easily described by an isotropic
quasi-Hookean \cite{cq:elastica,cs:superelastic} prescription. Then it
is simply assumed that $\Lambda_{\mrm{sol}} = -\muc s^2$ where
$\muc=\muc(n_\I)$ is the shear modulus and $s$ is a scalar measure of
the state of strain. Although this description is likely to be
adequate in any realistic situation we will nevertheless use the more
elaborate description discussed in Paper I and only later, in the
applications, specialise to isotropic quasi-Hookean solids. This will
make the resulting equations formally valid for anisotropic solids
without assuming the Hookean approximation. We will still use a
minimal coupling ansatz in the sense that we (quite reasonably) assume
that the solid contribution is independent of the free neutrons.  By
using the more general matter description, we have the advantage that
we can adopt the results of Paper IV more or less directly.

When describing solids it is important to keep track of the reference
state relative to which the strain is measured. For many simple solids
it is possible to do this via a positive definite metric tensor field,
the matter space metric $k_{ab}$ (see Paper I). One may intuitively
think of this tensor as encoding the (3-)geometry of the solid (as
seen by the solid itself) in an unstrained state.  The strain tensor
is then given by
\beq
  s_{ab} = \frac12(h_{ab} - n_\I^{-2/3}k_{ab})
\eeq
where $h_{ab}$ is defined in \refeq{nperp}.  It is advantageous to
work with an orthonormal eigenbasis $e_{\mu}^a$ of $k_{ab}$ in which
\beq
  k_{ab} = \sum_{\mu=1}^3 n_\mu^2 e_{\mu}^ae_{\mu}^b
\eeq
where we use Greek indices (with no implied sums) to enumerate the
basis and $n_\mu$ are, loosely speaking, linear particle densities. In
principle we could simply take the solidity contribution to the master
function to be a function of the $n_\mu$'s. For practical purposes,
however, it is convenient to instead work with the eigenvalues
$\alpha_\mu = \frac{n_\mu}{n^{1/3}_\I}$ of
\beq
  \eta_{ab} =  n_\I^{-2/3}k_{ab} 
           = \sum_{\mu=1}^3 \alpha_\mu^2 e_{\mu}^ae_{\mu}^b
\eeq
Since the determinant of the matter space metric is
given by $\det(k^a{}_{b})=n_\I^{2/3}$ it is clear that
\beq
  n_\I = \Pi_{\mu=1}^3 n_{\mu}
\eeq
It follows that $\eta^a{}_{b}$ has a unit determinant so that this
tensor only has two independent components. These components hold the
key information of the material's response to non-compressional
distortions and therefore encode the difference from a liquid. Within
the eigenvalue formulation it is natural to prescribe the solid
contribution as a function of the number density of confined baryons,
$n_\I$ and the parameters $\alpha_\mu$,
\beq
  \Lambda_{\mrm{sol}} = \Lambda_{\mrm{sol}}(n_\I, \alpha_\mu)
\eeq
making manifest the ``minimal coupling'' ansatz, \ie\ that we assume
that the solid's response to deformations is independent of the number
density of free neutrons.  In a quasi-Hookean approximation the
solid's function of state may be separated in the form
\beq
  \Lambda_{\mrm{sol}} = \muc(n_\I)s^2(\alpha_\mu)
\eeq
where $\muc$ is the shear modulus and $s^2$ is a scalar measure built
from invariants of the strain tensor (see Paper I for a discussion).

To summarise, the above prescription leads to a Lagrangian
that can be written
\beq
  \Lambda = \Lambda_{\mrm{liq}}(n_\I^a, n_\f^a) +
    \Lambda_{\mrm{sol}}(n_\I, \alpha_\mu)
\eeq
where the standard isotropic two-fluid Lagrangian takes the form
\beq
  \Lambda_{\mrm{liq}}  = -\rhoc + \frac{1}{n_\I}(m^*_\f - m)(n_{\I\f}^2-n_\I n_\f)
\eeq
Practically speaking, our matter model is described by three functions
of state; the minimum energy density $\rhoc$, the effective neutron
mass $m_\f^*$ and the solid's function of state which, in the
subsequent applications, will be encoded in the shear modulus $\muc$.

\subsection{Equations of motion}

The variational procedure described in CS, applied to the Lagrangian
density discussed above, leads to a stress-energy tensor of the form
\beq
  T^a{}_{b} = (\Lambda - n_\f^d\mu_d^\f -  n_\I^d\mu_d^\I)\delta^a{}_{b}
     +  n_\f^a\mu_b^\f + n_\I^a\mu_b^\I + \pi^a{}_b
\eeq
Here
\beq
  \mu_a^\f = \frac{\partial \Lambda}{\partial n^a_\f}
           = \frac{\partial \Lambda_{\mrm{liq}}}{\partial n^a_\f}, \qquad
  \mu_a^\I = \frac{\partial \Lambda}{\partial n^a_\I}
\eeq
are the momenta of the constituents and $\pi_{ab}$ is the (trace-less)
solid contribution as derived in Paper I. The stress energy tensor may
be used as source in Einstein's equations. In addition, we have two
equations of motion of the form
\begin{align}
  2n^a_\f\nabla_{[a}\mu^\f_{b]} &= 0 \lbeq{eulerf}\\
  2n^a_\I\nabla_{[a}\mu^\I_{b]} + \nabla^a\pi_{ab} &= 0 \lbeq{eulerc}
\end{align}
where square brackets indicate anti-symmetrisation. In the following,
we will use the fully variational description (CS). This means that
the constituents are individually conserved, \ie\ $\nabla_a n_\I^a =
\nabla_a n_\f^a = 0$. The equations \refeq{eulerf} and \refeq{eulerc}
(corresponding to the Euler equations) together imply the conservation
of energy momentum $\nabla^aT_{ab} = 0$. This is, of course, also
implied by the Einstein equations via the contracted Bianchi
identities. Thus the various equations are not independent. This
should be familiar from single fluid problem where it is well-known
that one can opt to work with the Einstein equations alone. For
multifluid problems more information is required. In the present
context it is sufficient to consider a combination of the Einstein
equations and one of the Euler equations. The simplest choice, since
we only \underline{need} to consider the complicated elastic terms
once, is to let \refeq{eulerf} do the job. As we will see, the
incompressible nature of the axial modes together with the fact that
the free neutrons are only forced via the entrainment then allows us
to find the linearised solutions to the mode problem explicitly. We
may also comment that the Cowling approximation, wherein the
gravitational degrees of freedom are neglected, retain the property
\cite{sa:axicowling} that it is equivalent to consider the weak
coupling limit of Einstein's equations together with
\refeq{eulerf}. Alternatively, one can decide to work only with
\refeq{eulerf} and \refeq{eulerc}. The latter strategy would lead to
a problem which is trivially related to the corresponding one in
Newtonian theory.

In order to obtain explicit equations we need to determine the
momenta. Following the notation of AC we write formally
\begin{align}
  \mu_a^\f &= \Bcal^\f n_a^\f + \Acal^{\I\f} n_a^\I \\
  \mu_a^\I &= \Bcal^\I n_a^\I + \Acal^{\I\f} n_a^\f
\end{align}
where
\begin{align}
  \Bcal^\f     &:= -2\frac{\partial \Lambda}{\partial n_\f^2}
                = 2\frac{\partial \rhoc}{\partial  n_\f^2}
                  + \frac{1}{n_\f}m_\I^\f + O(v^2)
                \approx \frac{1}{n_\f}(\rhoc' + m_\I^\f) \\
  \Bcal^\I     &:= -2\frac{\partial \Lambda}{\partial n_\I^2}
                =  2\frac{\partial \rhoc}{\partial  n_\I^2}
                  +\frac{n_\f}{n_\I^2}m_\I^\f
                  - 2\frac{\partial \Lambda_{{\mrm sol}}}{\partial n_\I^2} + O(v^2)
                \approx \frac{1}{n_\I}\left(\rhoc'
                      + \frac{n_\f}{n_\I}m_\I^\f\right)
                   - 2\frac{\partial \Lambda_{{\mrm sol}}}{\partial n_\I^2}  \\
  \Acal^{\I\f} &:= -\frac{\partial \Lambda}{\partial n_{\I\f}^2}
                = -\frac{1}{n_\I}m_\I^\f + O(v^2)
                \approx -\frac{1}{n_\I}m_\I^\f
\end{align}
Here we make use of the assumption that $\rhoc$ can be treated as a
function of $n$ only and denote its derivative with respect to $n$ by a
prime. The approximate expressions are valid up to  $O(v)$ which is
all we will need subsequently. We may note here that the generalised
pressure (see, \eg\ CS or AC) is
\beq
  \Psi := \Lambda - n_\I^a\mu_a^\I - n_\f^a\mu_a^\f
        = n\rhoc' - \rhoc  + \Lambda_{\mrm{sol}}
          - n_\I\frac{\partial \Lambda_{{\mrm sol}}}{\partial n_\I} + O(v^2)
        \approx \pc +  \Lambda_{\mrm{sol}}
          - n_\I\frac{\partial \Lambda_{{\mrm sol}}}{\partial n_\I}
\eeq
where $\pc$ denotes the usual (unsheared) pressure. Comparing to the
result in Paper I we see that $\Psi$ corresponds to the total
isotropic pressure defined there. In an unstrained state we have
\beq
  \rhoc' = \frac{\rhoc+\pc}{n} \approx \frac{\rhoc}{n} \approx m
\eeq
The approximations are valid in the relatively tenuous neutron star
crust where $\pc \ll \rhoc$ and $\rhoc \approx mn$. This
simplification, which is accurate to within $\sim 0.1$ \%, is
obviously very useful since we replace a function with a known
constant. Nevertheless, in our numerical code we use the full
expression (which actually means that we do not need to assume that
the neutrons and protons have the same mass).

In order to compare the present model to the purely elastic case
considered in Paper I it is useful to express the energy momentum
tensor in a frame adapted to the solid. Formally we then have
\beq
  T_{ab} = \rho u_au_b + 2u_{(a}Q_{b)} + P_{ab}
\eeq
where $\rho$, $Q_a$ and $P_{ab}$ are, respectively, the total energy
density, the momentum flow and the pressure tensor, all measured in a
frame described by the solids four-velocity $u^a$. Putting the pieces
together we find that
\begin{align}
  \rho &= \rho_I + O(v^2) \\
   Q_a &= x_\f(\rhoc + \pc)v_a + O(v^2) \lbeq{Q} \\
   P_{ab} &= \Psi h_{ab} + \pi_{ab} + O(v^2) = P^I_{ab} + O(v^2)
\end{align}
where $x_\f = n_\f/n$ is the free neutron fraction and we use the
label $I$ to denote the corresponding quantity in Paper I.  Thus, to
order $v$  the free neutrons lead to the presence of a
non-zero momentum flow $Q_a$. Obviously, a static background is
unaffected by this.

\section{Perturbations}
\label{sec:perturbations}

Let us now consider the equations governing axial perturbations around
a static background. We base our derivation of these equations on the
general framework developed by Karlovini \cite{karlovini:axial} and
make heavy use of the results in Paper IV to which we refer for
details.

The starting point of the analysis is a decomposition of the metric
in the form
\beq
  g_{ab} = \perp_{ab} + F^{-1}\eta_a\eta_b, \qquad
  F = \eta^a\eta_a = (r\sin\theta)^2, \qquad
  \eta^a\perp_{ab} = 0
\eeq
where $\eta^a$ is the axial Killing vector. Karlovini
\cite{karlovini:axial} showed that the axial part of the metric
perturbations (denoted here by $\gamma_{ab}$) can be entirely
expressed in terms of $\delta\eta_a$ so that $\delta F$ and
$\delta\!\!\perp_{ab}$ can both be set to zero\footnote{In order to
avoid introducing yet another $\mu$ we have chosen to express the
metric perturbations in terms of $\eta_a$ rather than $\mu_a =
F^{-1}\eta_a$ used in \cite{karlovini:axial} and Paper IV. Note that
$\delta\eta_a = F\delta\mu_a \neq g_{ab}\delta\eta^a = 0$ where the
latter equality is due to a partial gauge fixing (which is just for
convenience since the present formalism is gauge invariant).}. The
perturbed Einstein equations then take the simple form
\begin{align}
  \nabla_b(FQ^{ab}) &= \kappa J^a \\ \nabla_aJ^a &= 0
\end{align}
where, loosely speaking, $Q_{ab} = 2\nabla_{[a}F^{-1}\delta\eta_{b]}$
represent the geometric perturbations and
\beq\lbeq{Jdef}
  J^a = 2\delta(\perp^{ab}\eta^cT_{bc})
\eeq
encode the matter motion. Hence, it is natural to refer to $J^a$ as
the matter current. These equations were discussed in detail for the
case of a single solid in Paper IV. Here we focus on the changes
needed to extend those equations to the case where the solid is
coexisting with a superfluid. As noted above, the superfluid degrees
of freedom manifest themselves by i) the additional momentum flow
$Q^a$ in the energy momentum tensor, and ii) the need to consider the
equation of motion \refeq{eulerf}. In order to work out the explicit
perturbation equations we found it useful to make use of the
eigenvector formulation introduced in Paper I where the principal
directions of the solid provide an orthonormal tetrad $\{u^a,
e_\mu^a\}$, where $u^a$ is the four-velocity of the solid, $\mu \in
\{1,2,3\}$ and we use Greek indices to enumerate the spatial basis
vectors. The general expressions for the perturbed tetrad were given
in Paper IV in terms of the perturbed metrics $\gamma_{ab} = \delta
g_{ab}$ and $\delta k_{ab}$ which in the axial case considered here
can be written;
\beq
  \gamma_{ab} = 2F^{-1}\eta_{(a}\delta\eta_{b)}, \qquad
  \delta k_{ab} = 2n_3^2\eta_{(a}\nabla_{b)}\delta\tilde\phi
\eeq
The quantity $\tilde\phi$ is the (inverse) mapping of the azimuthal
coordinate on matter space (see Paper IV) and we use parentheses to
denote symmetrisation. In general, the inclusion of the free neutrons
adds four scalar degrees of freedom. We take these to be represented
by the free neutron density $n_\f$ and the relative velocity $v^a$
(which, due to the constraint $u^av_a=0$ has three degrees of
freedom). For the axial case it is straightforward to show that we
must have (for a static background)
\beq
  \delta n_\I = \delta n_\f = \delta n_{\I\f} = 0
\eeq
It follows that
\beq
  \delta \Bcal^\I = \delta \Bcal^\f = \delta \Acal^{\I\f} = 0
\eeq
so that the perturbed momenta are given by
\begin{align}
  \delta \mu_a^\I &= \Bcal^\I\delta n_a^\I + \Acal^{\I\f}\delta n_a^\f \\
  \delta \mu_a^\f &= \Bcal^\f\delta n_a^\f + \Acal^{\I\f}\delta n_a^\I
\end{align}
For the perturbations we shall consider the relative velocity to be
aligned with the axial Killing vector. Therefore we may write it in
the form
\beq
  \delta v^a = v^a = v\eta^a
\eeq
where we have dropped the perturbation symbol $\delta$ since no
confusion can arise.  This leads to a simple expression for the
perturbed neutron momentum,
\beq\lbeq{dmf}
  \delta\mu^\f_a = [(\rhoc'+m_c^f)v - \rhoc'u^bK_b]\eta_a
\eeq
where $K_a = \nabla_a\delta\tilde\phi - F^{-1}\delta\eta_a$ is a gauge
invariant one-form. As discussed in paper IV, it encodes the complete
nature of the perturbations of the solid. On a static background, and
only considering the non-stationary (oscillatory) perturbations of
equation \refeq{eulerf} we find the remarkably simple result
\beq
  \delta\mu^\f_a = 0
\eeq
This implies, via \refeq{dmf}, that we have an algebraic solution for
the perturbed relative velocity
\beq
  v = \frac{\rhoc'}{\rhoc' + m_c^f}u^aK_a \approx \frac{m}{m_\f^*}u^aK_a
\eeq
Note also that this shows that the vorticity remains zero to the first
order accuracy considered here. Thus, no vortices are created or
destroyed dynamically to this order. This simple result may seem
surprising at first, but it is a direct consequence of the
incompressible nature of axial perturbations. The variational
principle we use is based on varying the flowlines of the
particles. Incompressibility implies that the total number of
flowlines in a 3-volume is conserved and hence restricts the
variations further to the extent that only a single displacement
vector will be needed to describe the motion of the full system.

Armed with the solution to the equation of motion for the free
neutrons we are in a position to evaluate the effect the free neutrons
have on the perturbed energy momentum tensor. As already pointed out,
the only difference from the purely elastic case is the presence of
the momentum flow $Q_a$. Perturbing \refeq{Q} and inserting the result
in
\refeq{Jdef} we readily find
\beq
  J^a = 2(\rho+p_t)F\tilde{S}^{ab}K_b
\eeq
where $\rho$ and $p_t$ are the total energy density and the tangential
pressure, respectively, and the difference induced by the superfluid is
the modification of the shear wave ``metric'':
\beq
  \tilde{\Scal}^{ab} = \Scal^{ab} + x_\f\frac{\rhoc'}{\rhoc'+m_c^f}\frac{\rhoc+\pc}{\rho+p_t}u^au^b
\eeq
Here
\beq
  \Scal^{ab} = -u^au^b + \vr e_1{}^ae_1{}^b + \vt e_2{}^ae_2{}^b
\eeq
is the corresponding metric for purely elastic matter expressed in
terms of the eigenvector basis $e_\mu{}^a$ of the solid
lattice. Meanwhile, $v_{r\perp}$ and $v_{t\perp t}$ are the shear wave
velocities in the radial and tangential directions, respectively (see
Paper I). Hence, the effect of the neutrons is
\underline{entirely} contained in the factor
\beq
  \chi = 1 -  x_\f\frac{\rhoc'}{\rhoc'+m_c^f}\frac{\rhoc+\pc}{\rho+p_t}
\eeq
From Paper I we know that the change of the pressure and density due
to anisotropy is of the order $\sim\muc s^2$. In the bulk of the
crust we have the ordering $\muc\ll\pc\ll\rhoc$ and in addition 
$s^2\lesssim 0.1$ \cite{hk:breakingstrain}. Hence we expect that the
approximations $\rho \approx \rhoc$ and $p_t \approx
\pc$ will always hold in realistic neutron star crusts. They will, of
course, be exactly true on an unstrained background. We therefore note
that since $\rhoc'\approx m$ we have (to $\sim 0.1\%$ precision)
\beq
  \chi \approx 1 - x_\f\frac{m}{m_\f^*}
\eeq
in perfect agreement with result from a recent analysis of
incompressible plane waves in the corresponding Newtonian problem
\cite{ags:supsig}.  Thus, to ``upgrade'' the equations in Paper IV we
only need to insert the factor $\chi$ appropriately. The general
equations for a static background and $l \ge 2$ thus become
\begin{align}
  -\dot\Wcal_t + \Wcal_r' - \frac{r'}{r}\Wcal_r
       - e^{2\nu}\frac{L}{r^2}\psivar &= 0 \lbeq{1ogen1}\\
  -\dot{\Wcal}_r + \Wcal_t' +\frac{r'}{r}\Wcal_t
       + e^{2\nu}\frac{\Ecal\vt}{r^2} \varphi &= 0 \lbeq{1ogen2}\\
  -Lr\dot\psivar + \Ecal\vr r^2\Bp + (\Ecal\vr +L)r\Wcal_t &= 0  \lbeq{1ogen3}\\
  - \chi\Ecal r\dot\varphi + L\Ap - (\chi\Ecal+L)r\Wcal_r &= 0 \lbeq{1ogen4}
\end{align}
\cf\ equations (98)-(101) in Paper IV. Here $r$ is the radial measure
in Schwarzschild coordinates in which the line element takes the form
\beq
\d s^2 = -e^{2\nu}\d t^2 + e^{2\lambda}\d r^2
+ r^2(\d\theta^2 +
\sin^2\theta \d \phi^2)
\eeq
We are also using $L=(l-1)(l+2)$,
\beq
  \Ecal = 2\kappa r^2(\rho+p_t)
\eeq
and dots and primes refer to derivatives with respect to time and the
Regge-Wheeler radial coordinate $r_*$ which is given by
\beq
  \d r_* = e^{\lambda - \nu}\d r
\eeq
The total energy density $\rho$ and the tangential pressure $p_t$ (see
Paper I) have been retained to keep the formulae valid for anisotropic
backgrounds. We will later set them to the isotropic values $\rhoc$
and $\pc$, respectively. For an explanation of the dependent
variables, see Paper IV.  We see that only the last equation is
changed by the presence of the free neutrons.

For completeness we also give the modified wave equations (of which
only the second, describing the shear waves, is modified)
\begin{align}
  -\ddot\Wcal_t + \Wcal_t'' + e^{2\nu}\left(\frac{6m}{r^3} - \frac12\kappa(\rho-p_r)\right.
      & \left. -\frac{\Ecal\vr}{r^2} - \frac{l(l+1)}{r^2}\right)\Wcal_t \nonumber\\
      &= r\left[\frac{e^{2\nu}\Ecal(\vr-\vt)}{r^3}\varphi\right]'
        - \left[\frac{e^{2\nu}\Ecal\vr}{r^2}\right]'\varphi \lbeq{Weq1}\\
  -\chi\ddot\varphi + \frac{(\Ecal\vr\varphi')'}{\Ecal}
  - \left[\frac{(\Ecal\vr r')'}{\Ecal r}\right. & \left.+
      e^{2\nu}\frac{\vt(\chi\Ecal+L)}{r^2}\right]\varphi
      = \frac{[\Ecal(\chi-\vr)r\Wcal_t]'}{\Ecal r}
      - \frac{\chi\Ecal'}{\Ecal} \Wcal_t\lbeq{Weq2}
\end{align}
The boundary conditions, namely that $\Wcal_t$, $\Wcal_r$ and $\psi$
are everywhere continuous, do not change. Thus, we now have all the
information needed to solve the axial oscillation problem including
the free neutrons.

\section{Results}
\label{sec:results}

The basic strategy for solving the perturbation equations
\refeq{1ogen1} -- \refeq{1ogen4}, in the case where the free
neutrons are ignored, ($\chi=1$) has already been outlined in Paper
IV. Given that the free neutrons do not change the problem formally,
the same strategy can be used in the more general setting discussed
here. Nevertheless, it is worthwhile providing some detail on the
methods we have used to to solve the equations. Interested readers can
find a discussion of the relevant points in the Appendix.

We build our background neutron star models using realistic tabulated
equations of state, see table \ref{tab:bkgmodels}.
\begin{table}[t]
\begin{tabular}{l l c c c c c l}
\hline
Model & EoS & $\rhoc_0$ ($10^{15}$ g/cm$^3$) & $M (M_\odot)$ & $R$
(km) & $M_{\mrm{core}} (M_\odot)$ & $R_{\mrm{core}}$ (km) & $\beta$ \\
\hline
A1(a)   & A        & 1.259 & 1.04959 & 9.89188 & -       & -       & 0.156751 \tabnl
A1(b)   & A + DH   & 1.259 & 1.04806 & 9.90598 & 1.03690 & 8.99246 & 0.156300 \tabnl
A2(a)   & A        & 4.110 & 1.65241 & 8.37048 & -       & -       & 0.291632 \tabnl
A2(b)   & A + DH   & 4.110 & 1.65019 & 8.37450 & 1.64746 & 8.10619 & 0.291100 \tabnl
B1(a)   & B        & 1.995 & 0.97068 & 8.76793 & -       & -       & 0.163549 \tabnl
B1(b)   & B + DH   & 1.995 & 0.97045 & 8.78057 & 0.96302 & 8.01716 & 0.163274 \tabnl
B2(a)   & B        & 5.910 & 1.41131 & 7.07165 & -       & -       & 0.294828 \tabnl
B2(b)   & B + DH   & 5.910 & 1.41003 & 7.07679 & 1.40843 & 6.85589 & 0.294348 \tabnl
APR1(a) & APR      & 0.750 & 0.91989 & 11.5963 & -       & -       & 0.117189 \tabnl
APR1(b) & APR + DH & 0.750 & 0.92026 & 11.7203 & 0.89627 & 10.1619 & 0.115995 \tabnl
APR2(a) & APR      & 2.750 & 2.19428 & 10.0059 & -       & -       & 0.323969 \tabnl
APR2(b) & APR + DH & 2.750 & 2.19436 & 10.0238 & 2.19071 & 9.77753 & 0.323404 \\
\hline
\end{tabular}
\caption{Background models used in our study. We consider three equations 
of state, labelled A \cite{pandharipande:eosA}, B
\cite{pandharipande:eosB} and APR \cite{apr:eos}. The reason for this 
particular choice is simply that they were previously studied by BBF
\cite{bbf:wmodes}. This allows us to compare the results for the
$w$-modes directly. For each equation of state we consider the same
two central densities as BBF. For each of these models we also
substitute the equation of state by DH in the crust. As the data in
the table shows, this leads to slightly different total masses and
radii. For each model we provide the central density $\rho_0$, total
mass $M$, total radius $R$ and compactness $\beta=M/R$. For the models
where the crust-core transition pressure is known (\ie\ for the DH
crust) we also give the mass $M_{\mrm{core}}$ and radius
$R_{\mrm{core}}$ of the core.}
\label{tab:bkgmodels}
\end{table}
The two older equation of state tables ``A''
\cite{pandharipande:eosA} and ``B'' \cite{pandharipande:eosB} were taken
from the distribution of the {\texttt{rns}} code, see \eg\
\cite{sf:rns}, whereas  ``APR'' \cite{apr:eos} was provided by G.\ Ravenhall.
In the crust the equation of state is taken from Douchin \& Haensel
\cite{dh:eos} (DH)\footnote{In fact, in the outer crust we employ the
results described in \cite{samuelsson:thesis} using more recent
measurements for the binding energy of nuclei, but the end result is
practically indistinguishable from DH.}  and we use the shear modulus
for a Coulomb lattice as calculated by Ogata \& Ichimaru
\cite{oi:shearmod}\footnote{The effects of using the very recent results
of Horowitz \& Hughto \cite{hh:shearmod}, obtained from molecular dynamics
simulations, will be examined in a forthcoming paper.}.
The effective mass needed for the entrainment was
taken from Chamel \cite{chamel:mstarcrust}. We emphasise that, since his
calculations were performed for a different EoS than the ones we use,
our model for the effective mass is inconsistent. In addition, we
are only aware of data for four particular densities. Since we need data for
all crust densities, we use an analytic  expression that
approximately passes through the available data points, see
\cite{ags:supsig}. This is, obviously, not satisfactory but it is the
best that we can do at the present time.  Further work on the
properties of the crust ``beyond'' the EoS (minimum energy state)
should be encouraged. Even though our model is somewhat \emph{ad hoc},
it should provide insights into the basic dynamics of the problem.

As a first test of the numerical code we determined the $w$-modes for
a number of cases:

\noindent
(1a) using the fluid equations with the available tables, as described
above, all the way to the surface,

\noindent
(1b) as (1a) but substituting using the EoS of Douchin \& Haensel in
the crust region

\noindent
(2) artificially setting the shear modulus to zero,

\noindent
(3) using the full equations but artificially setting $\chi=1$, and
finally,

\noindent
(4) using the full equations including the model for the superfluid
neutrons.

Case (1a) can be directly compared to the results of Benhar \etal\
\cite{bbf:wmodes} (BBF). We compare a selection of typical results in
table \ref{tab:wmodes}.
\begin{table}[t]
\begin{tabular}{l l l l l l l}
\hline
Model & $f_{(a)}$ (kHz) & $f_{(b)}$ (kHz) & $f_\mathrm{BBF}$ (kHz) & $\tau_{(a)}$ ($\mu$s) & $\tau_{(b)}$ ($\mu$s) & $\tau_\mathrm{BBF}$ ($\mu$s) \\
\hline
A1   & 9.781128 & 9.785983 & 9.76 & 21.574378 & 21.540601 & 21.6 \tabnl
A2   & 9.107043 & 9.119139 & 9.11 & 72.028558 & 71.577816 & 72.4 \tabnl
B1   & 11.24895 & 11.25115 & 11.2 & 20.003988 & 19.999529 & 20.2 \tabnl
B2   & 10.65463 & 10.66564 & 10.6 & 70.601159 & 70.188532 & 71.7 \tabnl
APR1 & 8.865023 & 8.865084 & 8.82 & 19.195766 & 19.203608 & 19.4 \tabnl
APR2 & 6.725440 & 6.725176 & 6.69 & 158.52087 & 158.53555 & 165.3 \\
\hline
\end{tabular}
\caption{The data in this table compares the $w$-modes that we have 
determined to the results of BBF.  We give the frequencies and damping
times for the lowest $l=2$ $w$-modes for completely fluid stars. The
subscripts refer to the cases discussed in the text, \ie\ $(a)$ uses
the tables discussed in the main text and the fluid equations, $(b)$
substitutes the DH EoS in the crust and use the solid equations with
$\muc$ set to zero. Finally, BBF refers to the results in
\cite{bbf:wmodes}.  }
\label{tab:wmodes}
\end{table}
This comparison shows that the results are generally in good
agreement. The small differences could be due to slightly
different tabulations of the EoS or different interpolation schemes
(we use logarithmic interpolation). We also expect our treatment of
the surface of the background model to be more accurate. To rule out
the possibility that the difference is due to a bug in our code we
also calculated $w$-modes for a polytropic equation of state and
compared to the results of Andersson \& Kokkotas
\cite{ak:polytropes}. In this case (where the ambiguity in the
treatment of the EoS is eliminated) we find perfect agreement. A third
test of the code is provided by comparing cases (1b) and (2). The
agreement is excellent also in this case.

In order to investigate what effect the elastic solid and the
superfluid neutrons have on the $w$-modes it is relevant to compare
cases (2)--(4). Since the $w$-modes are of gravitational origin and
the matter motion is limited to the relatively tenuous crust we expect
a very small effect. The numerical results confirm this
expectation. We find that the relative influence of the crust and the
superfluid is less than $\sim 10^{-8}$ (see table \ref{tab:effects}
for some typical results). This difference is roughly at the same
level as the expected accuracy of our code. Thus, for all practical
purposes, the axial $w$-modes are unaffected by the crust properties.
\begin{table}[t]
\begin{tabular}{l l l}
\hline
Model & $f$ (kHz) & $\tau$ (ms) \\
\hline
Fluid      & 9.11913825  & 0.0715778249 \tabnl
Solid      & 9.11913834  & 0.0715778266 \tabnl
Solid + SF & 9.11913833  & 0.0715778264 \\
\hline
\end{tabular}
\caption{The data in this table illustrates the effects of the matter model.
We give the frequency and damping time of the lowest $l=2$ $w$-mode
for the background model A2(b) (see table \ref{tab:bkgmodels}) for
different matter models. The three models are: a completely fluid star
($\muc =0$, 'Fluid'), a star with a solid crust, but neglecting the
free neutrons ('Solid') and the complete model including the neutrons
('Solid +SF').  The results show that the effects due to the more
detailed treatment of the crust physics are tiny. The results have
converged (after increasing the resolution) to the stated precision,
but one should be aware of the fact that the roundoff error in the
calculation is, as discussed in the main text, of the order of $\sim
10^{-10}$.  }
\label{tab:effects}
\end{table}

Turning to the torsional $t$-modes, where one would expect the
superfluid to have a significant effect, we compare cases (2) and (3)
to our previous results \cite{sa:axicowling}, that were obtained
within the Cowling approximation. Since the estimated (via the
quadrupole formula) gravitational-wave damping time\footnote{Note that the estimates in
\cite{st:torsional} concern the damping time for the energy, while we
need the damping time associated with the amplitude. However, given
the order of magnitude nature of our discussion of this point we do
not bother with the different factors of 2 that relate these
quantities.} for the $t$-modes
is about $\tau \sim 10^4$ years \cite{st:torsional} we expect the real part of the
frequency to be well approximated by the Cowling results and the
imaginary part to be very small. Indeed, a typical $l=2$ fundamental
torsional mode has $f\sim 30$ Hz so that
\beq
  \Re(\omega) = 2\pi f \sim 1.9 \times 10^{2}\,\,\mbox{s}^{-1}
\eeq
In contrast, the imaginary part is
\beq
  \Im(\omega) = \frac{1}{\tau} \sim 0.8\times 10^{-12} \,\,\mbox{s}^{-1}
\eeq
some fourteen orders of magnitude smaller than the real part. Since
our numerical code uses an adaptive step Runge-Kutta integrator we can
control the accuracy in each step. However, using a small error
tolerance per step induces an uncontrollable truncation error (we use
double precision so that the truncation error is $\sim 10^{-15}$)
which, if turning down the tolerance on the error in each step,
requires more steps to complete the integration. A simple analysis,
using the \emph{de facto} used steps together with the prescribed
tolerance, suggests that our code cannot reach an accuracy beyond
about 1 part in $10^{10}$. Thus, we should not expect to be able to
determine the the damping time of the $t$-modes directly.  The
numerical calculations confirm this expectation.  When we feed the
output of the integrator into the M\"uller root-solver we find that we
cannot determine the imaginary part of the mode frequency. The real
part, on the other hand, converges nicely to more than ten digit
precision (while the imaginary part oscillates around zero). For this
reason we are confident that we determine the oscillation frequency
accurately. Typical results are provided in table \ref{tab:tmodes}. In
the table, we compare the results obtained using the Cowling
approximation (the case where the superfluid is neglected was
discussed in \cite{sa:axicowling} and the analysis of the problem
including the superfluid is currently being finalised).

\begin{table}[t]
\begin{tabular}{l c c c}
\hline
Model & ${}_2f_0$ (Hz) & ${}_2f_1$ (Hz) & ${}_2f_2$ (Hz) \\
\hline
A1(b)        & 31.097 (31.069) & 881.15 (881.15) & 1449.2 (1449.4) \tabnl
A1(b) + SF   & 33.897 (33.872) & 957.93 (957.92) & 1537.7 (1537.8) \tabnl
A2(b)        & 27.169 (27.160) & 1794.2 (1794.2) & 2962.9 (2963.3) \tabnl
A2(b) + SF   & 29.672 (29.665) & 1950.3 (1950.2) & 3164.7 (3165.1) \tabnl
B1(b)        & 34.549 (34.525) & 1031.5 (1031.5) & 1697.3 (1697.5) \tabnl
B1(b) + SF   & 37.666 (37.645) & 1121.4 (1121.4) & 1802.1 (1802.2) \tabnl
B2(b)        & 31.874 (31.867) & 2144.8 (2144.8) & 3542.1 (3542.5) \tabnl
B2(b) + SF   & 34.812 (34.806) & 2331.4 (2331.4) & 3783.6 (3784.0) \tabnl
APR1(b)      & 28.887 (28.841) & 584.00 (584.00) & 955.60 (955.66) \tabnl
APR1(b) + SF & 31.449 (31.408) & 634.90 (634.89) & 1008.0 (1008.1) \tabnl
APR2(b)      & 20.739 (20.731) & 1649.4 (1649.4) & 2724.7 (2725.0) \tabnl
APR2(b) + SF & 22.655 (22.647) & 1792.9 (1792.9) & 2912.1 (2912.5) \\
\hline
\end{tabular}
\caption{ This table provides a sample of results for the first few 
quadrupole $(l=2)$ $t$-modes. The background models are explained in
table \ref{tab:bkgmodels} and the extra label ``+ SF'' indicate that
we have taken the free neutrons into account. The results using the
Cowling approximation are shown within parenthesis for
comparison. Since the code we use for the Cowling approximation is
optimised for speed rather than accuracy it can only deliver 5-6 digit
precision which is why we only quote this accuracy here. The
calculations reported in this work is much higher, but this is most
likely irrelevant for astrophysical applications. We note that the
Cowling approximation in the cases tested is accurate to better than
about 0.01 \%. We also note that the case where the free neutrons are
taken into account typically gives $\sim$ 10 \% higher frequencies,
but we stress that this is model dependent and the effects can be much
larger (or smaller).}
\label{tab:tmodes}
\end{table}

\section{Discussion}
\label{sec:discussion}

In this paper we have considered the global axial quasi-normal modes
of relativistic stars with a (possibly) superfluid core and a solid
crust penetrated by a superfluid component. Our results provide the
first detailed analysis of this problem in general relativity, and
represent a key step towards the modelling of realistic neutron star
dynamics.  In fact, apart from the linear approximation common to all
mode studies and the omission of the magnetic field, our treatment
does not impose any significant approximations\footnote{The treatment
of the solid component assumes a conformally deforming solid, see
Paper I, but as this class incorporates, \eg\ cubic symmetric lattices
and isotropic solids we do not think that this constraint imposes any
important restrictions at the present time.}. The discussion does,
however, highlight the need for improved microphysics models. While we
have made an effort to use models that are as ``realistic'' as
possible, it is clear that the input parameters that we have used are
somewhat inconsistent. This is entirely due to the lack of complete
data from microphysics studies. Future tabulated equations of state
need to provide, in particular, the superfluid entrainment parameters
(both in the crust and in the core).  Once such data becomes
available, it will be straightforward to incorporate it in our
computational framework.  The issue concerning the neutron star
magnetic field is more challenging. This problem will require serious
thought, especially if we want to account for the presence of
superconducting protons in the core.

Specialising to an isotropic solid we determined both the axial
gravitational $w$-modes and the elastic torsional $t$-modes. In the
case of the $w$-modes we calculated both the frequency and
the damping time with high precision. The obtained results confirm the
expectation that these modes are very weakly influenced by the
presence of a solid/superfluid component. We also confirmed that the
frequencies of the $t$-modes are only weakly affected by the coupling
to the gravitational degrees of freedom. We were, however, unable to
directly determine the damping time for these modes. This is likely to
be completely irrelevant from an astrophysical point of view, since
other mechanisms will dominate the damping of these
modes. Of course, from an academic point of view the situation is not
entirely satisfactory. As a matter of principle, one would like to
have a direct determination of the damping times. Of course, a direct
application of the quadrupole formula should give reliable results. A
possible future option would be to take advantage of the enormous
difference between the real and imaginary parts of the quantities that
appear in the perturbation equations and perform a ``second
perturbation''. One can easily check that this leads to a set of
decoupled equations for the real parts and a set of equations for the
imaginary parts sourced by the real parts. We have not yet tried to
implement such a scheme.

As far as any magnetar seismology analysis is concerned
\cite{sa:axicowling}, the main lesson to learn from our analysis is that
one does not need an accurate treatment of the gravitational degrees
of freedom. However, the presence of the superfluid component in the
crust is important. We are currently investigating this problem in
more detail within the relativistic Cowling approximation, and hope to
be able to report on the results in the near future.  Having developed
the framework for the axial perturbations for relativistic neutron
star models with elastic and superfluid components, we also need to
consider the (generally more complex) problem of polar
perturbations. Developments in this direction are also under way.

\section*{Acknowledgements}
This work was supported by STFC through grant number PP/E001025/1. We
also acknowledge support from the EU-network ILIAS, providing
opportunity for valuable discussions with our European colleagues. We
like to express gratitude to Nicolas Chamel and Kostas Glampedakis for
valuable discussions and to Geoff Ravenhall for providing the tabulated
version of the APR EoS.

\appendix
\section[A]{Appendix: Numerical formalism}
\label{sec:numerics}

In this appendix we discuss the methods we used to
solve the axial perturbation  equations \refeq{1ogen1} -- \refeq{1ogen4}.
In paper IV the problem was analysed for the case $\chi=1$, \ie\ when the
free neutrons are ignored. Given that there is no real formal difference
between the two systems that need to be solved, we can use the
same strategy also for this more general problem.

In the interior of the star we use a standard
Runge-Kutta ODE solver from the GSL libraries \cite{getal:gsl}.   The
exterior perturbations are determined by a pseudo-spectral method (see
\cite{sam:characteristic} and below).

\subsection{Background solution}

A static, spherically symmetric, unstrained solution does not depend
on the elastic/superfluid nature of the matter content since both the
strain and entrainment contributions vanish. Thus, the background
solution is, in this case, identical to that of a perfect fluid. It is
well known that the use of a radial coordinate as independent
variable, as in the standard TOV equations, leads to numerical
difficulties near the surface. One problem is the steep gradient of
the fluid variables near the surface and another is that the surface
itself is determined by the condition that the radial pressure
vanishes. That is, it is determined by one of the dependent
variables. Lindblom
\cite{lindblom:phase} suggested that a better strategy, as far as the
background is concerned, is to define the two dependent variables
\beq
  u = r^2, \qquad v=\frac{m}{r}
\eeq
and use the relativistic enthalpy $h$ defined by
\beq
  \d h = \frac{\d \pc}{\rhoc + \pc}
\eeq
as the independent variable. Then the equations of structure take the
form
\begin{align}
  \frac{\d u}{\d h} &= -\frac{4u(1-2v)}{\kappa \pc u + 2v} =: u' \\
  \frac{\d v}{\d h} &= -(1-2v)\frac{\kappa\rhoc u - 2v}{\kappa \pc u + 2v}
                     = \frac{u'}{4u}(\kappa \rhoc u - 2v)
\end{align}
In order to ensure a regular centre the integration is started at
finite radius using the expansion
\begin{align}
  u &= \frac{12(h_0-h)}{\kappa(\rhoc_0+3\pc_0)} + \ldots \\
  v &= \frac{2\rhoc_0(h_0-h)}{\rhoc_0+3\pc_0} + \ldots
\end{align}
where a subscript '$0$' will be used throughout our discussion to denote evaluation
at the centre.

\subsection{Perturbations}

As we are interested in quasi-normal modes we make the standard
assumption that the time dependence is given by $\exp(i\omega t)$
where $\omega$ is a complex constant whose real and imaginary parts
represent the angular frequency and the damping timescale, respectively. Inside the
star we solve the perturbation equations together with the background
equations.

\subsubsection{The vacuum region}

In the surrounding vacuum region the axial perturbation equations reduce to
the standard Regge-Wheeler equation. For quasi-normal modes we require that
the solutions are outgoing waves at null infinity. This problem was
discussed in detail by Samuelsson, Andersson and Maniopoulou
\cite{sam:characteristic} and we apply their code as it is. Then,
given the mass $M$, radius $R$ and angular frequency $\omega$ we obtain the surface
value $g_s$ of a certain phase function $g$,
\beq \lbeq{phaseg}
  g_s = g_{|r=R} =\left.\left(\frac{1}{\psi}\frac{d\psi}{d r}\right)\right|_{r=R}
         + \frac{i\omega R}{R-2M}
\eeq

\subsubsection{The fluid core}

Following Paper IV we adapt the independent variables to the centre of
the star by defining
\begin{align}
  \Xcal_1 &= r^{-l-1}\Wcal_t\\
  \Xcal_2 &= -i\omega e^{h_0-\nu_s}r^{-l}\Wcal_r \\
  \Xcal_3 &= -i\omega e^{h_0-h}r^{-l-1}\psivar \\
  \Xcal_4 &= \omega^2 e^{2h_0-h-\nu_s}r^{-l}\varphi
\end{align}
where $\nu_s = \frac12\ln\left(1 - \frac{2M}{R}\right)$ denotes the
surface value of $\nu$.
In the fluid region the equations then reduce to
\begin{align}
  \frac{\d\Xcal_1}{\d h}
     &= -\frac{u'}{2u} \left[(l+2)\Xcal_1 + \frac{e^{h-h_0}}{\sqrt{1-2v}}\Xcal_2\right] \\
  \frac{\d\Xcal_2}{\d h}
     &= -\frac{u'}{2u}\left\{\frac{e^{h-h_0}}{\sqrt{1-2v}}\left[(l+2)(l-1)e^{2(h_0-h)}
       - \omega^2e^{2(h_0-\nu_s)}u\right]\Xcal_1 + (l-1)\Xcal_2\right\}
\end{align}
and the constraints
\beq\lbeq{pfconst}
  \Xcal_3= -e^{h_0-h}\Xcal_1, \qquad \Xcal_4 = -e^{h_0-h}\Xcal_2
\eeq
We may note here that the perturbations in the core do not depend on
the multifluid nature of the medium (\ie\ $\chi$ does not
appear). This is due to the fact that, like an ordinary perfect fluid,
this type of matter cannot sustain axial oscillations, but only
stationary currents. Hence, the only oscillations that remain are those
associated with the gravitational degrees of freedom (the
$w$-modes). This means that, one cannot use   the axial
$w$-modes to distinguish between non-rotating single- and multi-fluid stars.
If the star is rotating, however, it is known that the
``axial-led'' inertial modes will depend on the superfluid nature of matter
\cite{pha:superrot}.

In order to make sure that we have a regular solution near the origin
we expand the solution according to
\begin{align}
  \Xcal_1 &= \hat\Xcal_1 \left(1 -  \frac{6e^{2(h_0-\nu_s)}\omega^2 + \kappa(l+2)[3\pc_0 - (2l-1)\rhoc_0]}
      {(2l+3)\kappa(\rhoc_0+3\pc_0)}(h_0-h) + \ldots\right) \\
  \Xcal_2 &= \hat\Xcal_1 \left(-(l+2) + \frac{6(l+4)e^{2(h_0-\nu_s)}\omega^2 - (l+2)(l-1)\kappa[3\pc_0
       + (2l+7)\rhoc_0]}{(2l+3)\kappa(\rhoc_0+3p_0)}(h_0-h) + \ldots\right)
\end{align}
where $\hat\Xcal_1$ represent the arbitrary scaling of the solutions
and one should keep in mind that we are allowed to rescale the
solutions at our convenience. We check that the solution does not
depend on the (small) value chosen for $h_0-h$.

\subsubsection{The crust}

In the crust we have chosen to integrate a slightly different set of
equations with dependent variables given by
\begin{align}
 \Ycal_1 &= r\Wcal_t \\
 \Ycal_2 &= \frac1{i\omega r}\Wcal_r \\
 \Ycal_3 &= r(\Wcal_t-i\omega\psi) \\
 \Ycal_4 &= \frac1{r}\varphi + \frac1{i\omega r}\Wcal_r
\end{align}
For these variables $\Ycal_{3-4}$ are constrained to vanish in a fluid
and $\Ycal_3=0$ also in vacuum. Moreover, $\Ycal_{1-3}$ are everywhere
continuous. Since we lack concrete information about any anisotropy in
the solid we specialise the equations to the isotropic case where
\beq
  \vr = \vt = \frac{\muc}{\rhoc+\pc}, \qquad \rho = \rhoc, \qquad p_t = p_r = \pc
\eeq
The set of equations we solve then becomes
\begin{align}
  {\Ycal_1}_{,h} &= -Pu\left[\omega^2 \Ycal_2
                    - 2\kappa e^{2(\nu_s-h)}\muc(\Ycal_2-\Ycal_4)\right]\\
  {\Ycal_2}_{,h} &= \frac{P}{u^2\omega^2}\left[\omega^2u\Ycal_1
                    - Le^{2(\nu_s-h)}(\Ycal_1-\Ycal_3)\right]\\
  {\Ycal_3}_{,h} &= \frac{2\kappa Pu}{L}\left[\chi\omega^2u(\rhoc+\pc)\Ycal_4
                   + Le^{2(\nu_s-h)}\muc(\Ycal_2-\Ycal_4)\right]\\
  {\Ycal_4}_{,h} &= -\frac{LP}{2\kappa u^2\omega^2\muc}\left[\omega^2\Ycal_3
                   + 2\kappa e^{2(\nu_s-h)}\muc(\Ycal_1 - \Ycal_3)\right] \lbeq{eY4}
\end{align}
where $P = u'e^{\nu_s-h}/2\sqrt{u(1-2v)}$.

\subsubsection{Boundary conditions}

As outlined above, the boundary conditions both at the centre of the
star and at infinity are already taken care of in our scheme. It thus
remains to ensure that the proper jump conditions are satisfied at the
top and bottom of the crust. At these interfaces we must ensure that
$\Ycal_{1-3}$ are continuous. As discussed in Paper IV, we have no
local conditions on the fourth variable. However, a generic choice of
$\Ycal_4$ at the inner (say) boundary will not admit a continuous
solution at the outer boundary. For this reason we solve for two
linearly independent solutions in order to find the unique\footnote{
The solution will be unique only if one assumes that $\Ycal_4$ is
continuous in the crust. This would seem natural, but is in fact not
required by the equations. One could imagine discontinuities at \eg\
phase transitions between different regions in the solid. We take
$\Ycal_4$ to be continuous, corresponding to the assumption that the
crust behaves as a single solid. It is our understanding that the
sharp discontinuities found in calculations of nuclear matter at
absolute zero temperature are likely to be smoothed in a real neutron
star. At least on the length-scales of interest in a global mode
analysis. Moreover we find it highly unlikely that, even if such
discontinuities appear, the layers will slip freely along the
boundaries.} linear combination that satisfies all boundary
conditions.

There is one difficulty with this approach, however. The first term in
equation \refeq{eY4} is proportional to $\Ycal_3/\muc$. Near the
surface of the star the boundary conditions require that
$\Ycal_3\rightarrow 0$ which balances the smallness of the shear
modulus $\muc$ in that region. Unfortunately, neither of the two
outgoing solutions that we compute would typically satisfy the
boundary conditions at the surface (only a specific linear combination
will). Hence, this term will typically become very large close to the
surface which makes it difficult to find an accurate solution. To
remedy this situation we integrate the eigenfunctions in the crust
both from the top and the bottom to some intermediate matching
point. We check that the choice of matching point does not affect the
solutions.

The explicit matching conditions are as follows. At the fluid-solid
interface we have
\begin{align}
  \Ycal_1 &= r^{(l+2)}\Xcal_1 \\
  \Ycal_2 &= e^{\nu_c}r^{(l-1)}\omega^{-2} \Xcal_2\\
  \Ycal_3 &= 0 \\
  \Ycal_4 &= \mbox{free}
\end{align}
where the $\Xcal_i$'s are given by the fluid solution.  We define the
two linearly independent solutions with the interface values
\beq
  \Ycal^{(1)}_i = [r^{(l+2)}\Xcal_1, e^{\nu_c}r^{(l-1)}\omega^{-2} \Xcal_2, 0, 0],
  \qquad \Ycal^{(2)}_i = [0,0,0,1]
\eeq
which guarantees (using the freedom to rescale the fluid
solution) that the linear combination $\Ycal^{(\mrm{tot})}_i =
\Ccal_1\Ycal^{(1)}_i + \Ccal_2\Ycal^{(2)}_i$ satisfies the boundary conditions
at the surface for any values of $\Ccal_i$. At the surface we
arbitrarily fix the overall normalisation of the solution by setting
$\tilde\Ycal_1 = 1$ where we will use a tilde to distinguish the outer
crust solution from the inner. From the vacuum value of the phase
function \refeq{phaseg} we then obtain the boundary value of
$\tilde\Ycal_2$,
\beq
  \tilde\Ycal_2 = \frac{i\omega R - (R-2M)(R^{-1}+g_s)}{R^3\omega^2}
\eeq
In addition we require that $\tilde\Ycal_3=0$. Thus, we choose the independent
solutions via the interface values
\beq
  \tilde\Ycal^{(1)}_i = [1, \tilde\Ycal_2(g_s), 0, 0],
  \qquad \tilde\Ycal^{(2)}_i = [0,0,0,1]
\eeq
This again guarantees that the linear combination $\tilde\Ycal^{(\mrm{tot})}_i =
\tilde\Ycal^{(1)}_i + \Ccal_3\tilde\Ycal^{(2)}_i$ satisfies the boundary conditions
at the interface for any value of $\Ccal_3$. Note that we only have a
single coefficient in the linear combination due to the choice of
normalisation.

We have now ensured that the boundary conditions are fulfilled
everywhere except at the matching point. Here we require that all
$\Ycal_i$ are continuous. Thus we demand that
\beq
  \Ccal_1\Ycal^{(1)}_i + \Ccal_2\Ycal^{(2)}_i = \tilde\Ycal^{(1)}_i + \Ccal_3\tilde\Ycal^{(2)}_i
\eeq
This condition can be put in matrix form as
\beq
  {\mathbf{M}}\bar{v} = 0
\eeq
where
\begin{displaymath}
{\mathbf{M}} = \left(\begin{array}{cccc}
    \Ycal^{(1)}_1 & \Ycal^{(2)}_1 & \tilde\Ycal^{(1)}_1 & \tilde\Ycal^{(2)}_1 \\
    \Ycal^{(1)}_2 & \Ycal^{(2)}_2 & \tilde\Ycal^{(1)}_2 & \tilde\Ycal^{(2)}_2 \\
    \Ycal^{(1)}_3 & \Ycal^{(2)}_3 & \tilde\Ycal^{(1)}_3 & \tilde\Ycal^{(2)}_3 \\
    \Ycal^{(1)}_4 & \Ycal^{(2)}_4 & \tilde\Ycal^{(1)}_4 & \tilde\Ycal^{(2)}_4 \\
  \end{array}\right)
\end{displaymath}
and
\beq
 \bar v = [\Ccal_1, \Ccal_2, -1, -\Ccal_3]^T
\eeq
In order for a solution exist we need
$\det(\mathbf{M})=\det[\mathbf{M}(\omega)]=0$ which is the eigenvalue
equation that gives the quasi-normal frequencies. We solve the mode
condition using a standard complex root finder based on M\"uller's method.


\begin{thebibliography}{10}

\bibitem{sw:qpo}
T.~E. {Strohmayer} and A.~L. {Watts}.
\newblock {Discovery of Fast X-Ray Oscillations during the 1998 Giant Flare
  from SGR 1900+14}.
\newblock {\em \apjl}, 632:L111--L114, October 2005.
\newblock \eprint{arXiv:astro-ph/0508206}.

\bibitem{sw:flare2}
A.~L. {Watts} and T.~E. {Strohmayer}.
\newblock {Detection with RHESSI of High-Frequency X-Ray Oscillations in the
  Tail of the 2004 Hyperflare from SGR 1806-20}.
\newblock {\em \apjl}, 637:L117--L120, February 2006.
\newblock \eprint{arXiv:astro-ph/0512630}.

\bibitem{israel:qpo}
G.~L. {Israel}, T.~{Belloni}, L.~{Stella}, Y.~{Rephaeli}, D.~E. {Gruber},
  P.~{Casella}, S.~{Dall'Osso}, N.~{Rea}, M.~{Persic}, and R.~E. {Rothschild}.
\newblock {The Discovery of Rapid X-Ray Oscillations in the Tail of the SGR
  1806-20 Hyperflare}.
\newblock {\em \apjl}, 628:L53--L56, July 2005.
\newblock \eprint{arXiv:astro-ph/0505255}.

\bibitem{dt:magnetars}
R.~C. {Duncan} and C.~{Thompson}.
\newblock {Formation of very strongly magnetized neutron stars - Implications
  for gamma-ray bursts}.
\newblock {\em \apjl}, 392:L9--L13, June 1992.

\bibitem{duncan:1998A}
R.~C. {Duncan}.
\newblock {Global Seismic Oscillations in Soft Gamma Repeaters}.
\newblock {\em \apjl}, 498:L45--L49, May 1998.
\newblock \eprint{arXiv:astro-ph/9803060}.

\bibitem{sa:axicowling}
L.~{Samuelsson} and N.~{Andersson}.
\newblock {Neutron star asteroseismology. Axial crust oscillations in the
  Cowling approximation}.
\newblock {\em \mnras}, 374:256--268, January 2007.
\newblock \eprint{arXiv:astro-ph/0609265}.

\bibitem{levin:magnetars}
Y.~{Levin}.
\newblock {QPOs during magnetar flares are not driven by mechanical normal
  modes of the crust}.
\newblock {\em \mnras}, 368:L35--L38, May 2006.
\newblock \eprint{arXiv:astro-ph/0601020}.

\bibitem{gsa:mhd}
K.~{Glampedakis}, L.~{Samuelsson}, and N.~{Andersson}.
\newblock {Elastic or magnetic? A toy model for global magnetar oscillations
  with implications for quasi-periodic oscillations during flares}.
\newblock {\em \mnras}, 371:L74--L77, September 2006.
\newblock \eprint{arXiv:astro-ph/0605461}.

\bibitem{piro:flares}
A.~L. {Piro}.
\newblock {Shear Waves and Giant-Flare Oscillations from Soft Gamma-Ray
  Repeaters}.
\newblock {\em \apjl}, 634:L153--L156, December 2005.
\newblock \eprint{arXiv:astro-ph/0510578}.

\bibitem{sks:torsional}
H.~{Sotani}, K.~D. {Kokkotas}, and N.~{Stergioulas}.
\newblock {Torsional oscillations of relativistic stars with dipole magnetic
  fields}.
\newblock {\em \mnras}, 375:261--277, February 2007.
\newblock \eprint{arXiv:astro-ph/0608626}.

\bibitem{sksv:torsionalII}
H.~{Sotani}, K.~D. {Kokkotas}, N.~{Stergioulas}, and M.~{Vavoulidis}.
\newblock {Torsional Oscillations of Relativistic Stars with Dipole Magnetic
  Fields II. Global Alfv\'en Modes}.
\newblock {\em ArXiv Astrophysics e-prints}, November 2006.
\newblock \eprint{arXiv:astro-ph/0611666}.

\bibitem{cbk:QPOs}
A.~{Colaiuda}, H.~{Beyer}, and K.~D. {Kokkotas}.
\newblock {On the Quasi-Periodic Oscillations of Magnetars}.
\newblock {\em ArXiv e-prints}, February 2009.
\newblock \eprint{arXiv:0902.1401}.

\bibitem{csf:QPOs}
P.~{Cerd{\'a}-Dur{\'a}n}, N.~{Stergioulas}, and J.~A. {Font}.
\newblock {Alfven QPOs in magnetars in the anelastic approximation}.
\newblock {\em ArXiv e-prints}, February 2009.
\newblock \eprint{arXiv:0902.1472}.

\bibitem{levin:magnetarsII}
Y.~{Levin}.
\newblock {On the theory of magnetar QPOs}.
\newblock {\em \mnras}, 377:159--167, May 2007.
\newblock \eprint{arXiv:astro-ph/0612725}.

\bibitem{lee:aximag}
U.~{Lee}.
\newblock {Axisymmetric oscillations of magnetic neutron stars}.
\newblock {\em \mnras}, 374:1015--1029, January 2007.
\newblock \eprint{arXiv:astro-ph/0610182}.

\bibitem{lee:aximagII}
U.~{Lee}.
\newblock {Axisymmetric toroidal modes of magnetized neutron stars}.
\newblock {\em \mnras}, 385:2069--2079, April 2008.
\newblock \eprint{arXiv:0710.4986}.

\bibitem{ags:supsig}
N.~{Andersson}, K.~{Glampedakis}, and L.~{Samuelsson}.
\newblock {Superfluid signatures in magnetar seismology}.
\newblock {\em \mnras}, December 2008.
\newblock \eprint{arXiv:0812.2417}.

\bibitem{ks:qnm}
Kostas~D. Kokkotas and Bernd~G. Schmidt.
\newblock Quasi-normal modes of stars and black holes.
\newblock {\em \lr}, 2:2, 1999.
\newblock \eprint{arXiv:qr-qc/9909058}.

\bibitem{st:torsional}
Bonny~L. Schumaker and Kip~S. Thorne.
\newblock Torsional modes of neutron stars.
\newblock {\em \mnras}, 203:457, 1983.

\bibitem{cs:superelastic}
B.~{Carter} and L.~{Samuelsson}.
\newblock {Relativistic mechanics of neutron superfluid in (magneto) elastic
  star crust}.
\newblock {\em \cqg}, 23:5367--5388, August 2006.
\newblock \eprint{arXiv:gr-qc/0605024}.

\bibitem{ks:relasticityI}
Max Karlovini and Lars Samuelsson.
\newblock Elastic stars in general relativity: I. {F}oundations and equilibrium
  models.
\newblock {\em \cqg}, 20:3613--3648, 2003.
\newblock \eprint{arXiv:gr-qc/0211026}.

\bibitem{ksz:stability}
Max Karlovini, Lars Samuelsson, and Moundheur Zarroug.
\newblock Elastic stars in general relativity: {II}. {R}adial perturbations.
\newblock {\em \cqg}, 21:1559--1581, 2004.
\newblock \eprint{arXiv:gr-qc/0309056}.

\bibitem{ks:exact}
Max Karlovini and Lars Samuelsson.
\newblock Elastic stars in general relativity {III}. {S}tiff ultrarigid exact
  solutions.
\newblock {\em \cqg}, 21:4531--4548, 2004.
\newblock \eprint{arXiv:gr-qc/0401115}.

\bibitem{ks:relastaxial}
Max Karlovini and Lars Samuelsson.
\newblock Elastic stars in general relativity: {IV}. {A}xial perturbations.
\newblock {\em \cqg}, 24:3171--3189, 2007.
\newblock \eprint{arXiv:gr-qc/0703001}.

\bibitem{ac:sfreview}
Nils Andersson and Gregory~L. Comer.
\newblock Relativistic fluid dynamics: Physics for many different scales.
\newblock {\em Living Reviews in Relativity}, 10(1), 2007.
\newblock \eprint{arXiv:gr-qc/0605010}.

\bibitem{wald:gr}
Robert~M. Wald.
\newblock {\em General Relativity}.
\newblock The University of Chicago Press, Chicago, 1984.

\bibitem{cj:entrainment}
G.~L. {Comer} and R.~{Joynt}.
\newblock {Relativistic mean field model for entrainment in general
  relativistic superfluid neutron stars}.
\newblock {\em \prd}, 68(2):023002, July 2003.
\newblock \eprint{arXiv:gr-qc/0212083}.

\bibitem{chamel:2fluid}
N.~{Chamel}.
\newblock {Two-fluid models of superfluid neutron star cores}.
\newblock {\em \mnras}, 388:737--752, August 2008.
\newblock \eprint{arXiv:0805.1007}.

\bibitem{cq:elastica}
B.~Carter and H.~Quintana.
\newblock Foundations of general relativistic high pressure elasticity theory.
\newblock {\em \prsla}, 331:57, 1972.

\bibitem{karlovini:axial}
Max Karlovini.
\newblock Axial perturbations of general spherically symmetric spacetimes.
\newblock {\em \cqg}, 19:2125--2140, 2002.
\newblock \eprint{arXiv:gr-qc/0111066}.

\bibitem{hk:breakingstrain}
C.~J. {Horowitz} and K.~{Kadau}.
\newblock Private communication.

\bibitem{pandharipande:eosA}
V.~R. {Pandharipande}.
\newblock {Dense neutron matter with realistic interactions}.
\newblock {\em Nuclear Physics A}, 174:641--656, October 1971.

\bibitem{pandharipande:eosB}
V.~R. {Pandharipande}.
\newblock {Hyperonic matter}.
\newblock {\em Nuclear Physics A}, 178:123--144, December 1971.

\bibitem{apr:eos}
A.~Akmal, V.~R. Pandharipande, and D.~G. Ravenhall.
\newblock Equation of state of nucleon matter and neutron star stucture.
\newblock {\em \prc}, 58(3):1804--1828, September 1998.
\newblock \eprint{arXiv:hep-ph/9804388}.

\bibitem{bbf:wmodes}
O.~{Benhar}, E.~{Berti}, and V.~{Ferrari}.
\newblock {The imprint of the equation of state on the axial w-modes of
  oscillating neutron stars}.
\newblock {\em \mnras}, 310:797--803, December 1999.
\newblock \eprint{arXiv:gr-qc/9901037}.

\bibitem{sf:rns}
N.~{Stergioulas} and J.~L. {Friedman}.
\newblock {Comparing models of rapidly rotating relativistic stars constructed
  by two numerical methods}.
\newblock {\em \apj}, 444:306--311, May 1995.
\newblock \eprint{arXiv:astro-ph/9411032}.

\bibitem{dh:eos}
F.~{Douchin} and P.~{Haensel}.
\newblock {A unified equation of state of dense matter and neutron star
  structure}.
\newblock {\em \aaa}, 380:151--167, December 2001.
\newblock \eprint{arXiv:astro-ph/0111092}.

\bibitem{samuelsson:thesis}
Lars Samuelsson.
\newblock {\em Stellar Models in General Relativity}.
\newblock PhD thesis, Stockholm University, 2003.
\newblock ISBN 9172657464.

\bibitem{oi:shearmod}
S.~{Ogata} and S.~{Ichimaru}.
\newblock {First-principles calculations of shear moduli for Monte
  Carlo-simulated Coulomb solids}.
\newblock {\em \pra}, 42:4867--4870, October 1990.

\bibitem{hh:shearmod}
C.~J. {Horowitz} and J.~{Hughto}.
\newblock {Molecular Dynamics Simulation of Shear Moduli for Coulomb Crystals}.
\newblock {\em ArXiv e-prints}, December 2008.
\newblock \eprint{arXiv:0812.2650}.

\bibitem{chamel:mstarcrust}
N.~{Chamel}.
\newblock {Effective mass of free neutrons in neutron star crust}.
\newblock {\em Nuclear Physics A}, 773:263--278, July 2006.
\newblock \eprint{arXiv:nucl-th/0512034}.

\bibitem{ak:polytropes}
N.~{Andersson} and K.~D. {Kokkotas}.
\newblock {Pulsation modes for increasingly relativistic polytropes}.
\newblock {\em \mnras}, 297:493--496, June 1998.
\newblock \eprint{arXiv:gr-qc/9706010}.

\bibitem{getal:gsl}
Mark Galassi, Jim Davies, James Theiler, Brian Gough, Gerard Jungman, Michael
  Booth, and Fabrice Rossi.
\newblock {\em Gnu Scientific Library Reference Manual}.
\newblock Revised second edition.
\newblock ISBN 0954161734.

\bibitem{sam:characteristic}
L.~{Samuelsson}, N.~{Andersson}, and A.~{Maniopoulou}.
\newblock {A characteristic approach to the quasi-normal mode problem}.
\newblock {\em Classical and Quantum Gravity}, 24:4147--4160, August 2007.
\newblock \eprint{arXiv:0705.4585}.

\bibitem{lindblom:phase}
Lee Lindblom.
\newblock Phase transitions and the mass-radius curves of relativistic stars.
\newblock {\em \prd}, 58:024008, 1998.
\newblock \eprint{arXiv:gr-qc/9802072}.

\bibitem{pha:superrot}
A.~{Passamonti}, B.~{Haskell}, and N.~{Andersson}.
\newblock {Oscillations of Rapidly Rotating Superfluid Stars}.
\newblock {\em ArXiv e-prints}, December 2008.
\newblock \eprint{arXiv:0812.3569}.

\end{thebibliography}

\end{document}